\documentclass[aps,prl,twocolumn,superscriptaddress]{revtex4-1}

\usepackage{graphicx}

\usepackage{amsmath}
\usepackage{amssymb}
\usepackage{graphicx}\usepackage{afterpage}
\usepackage{amsfonts}
\usepackage{amssymb}
\usepackage{graphicx}
\usepackage{braket}
\usepackage{textcomp}
\usepackage{color}

\usepackage{xr}
\newcommand{\Ea}{\ensuremath{{\cal E}_1}}
\newcommand{\Eb}{\ensuremath{{\cal E}_2}}
\newcommand{\Ec}{\ensuremath{{\cal E}_3}}

\newcommand{\Ed}{\ensuremath{{\cal E}_{1,2}}}
\newcommand{\Er}{\ensuremath{{\cal E}_{\rm R}}}
\newcommand{\Eo}{\ensuremath{{\cal E}_{1,2,3}}}

\renewcommand{\figurename}{Figure}

\begin{document}
\title{Impact of environment on dynamics of exciton complexes in a WS$_2$ monolayer}

\author{T.~Jakubczyk}
\email[]{tomasz.jakubczyk@neel.cnrs.fr} \affiliation{Univ. Grenoble
Alpes, F-38000 Grenoble, France} \affiliation{CNRS, Institut
N\'{e}el, "Nanophysique et semiconducteurs" group, F-38000 Grenoble,
France}

\author{K.~Nogajewski}
\affiliation{Laboratoire National des Champs Magn\'{e}tiques
Intenses, CNRS-UGA-UPS-INSA-EMFL, 25 Av. des Martyrs, 38042
Grenoble, France}

\author{M.~R.~Molas}
\affiliation{Laboratoire National des Champs Magn\'{e}tiques
Intenses, CNRS-UGA-UPS-INSA-EMFL, 25 Av. des Martyrs, 38042
Grenoble, France}

\author{M.~Bartos}
\affiliation{Laboratoire National des Champs Magn\'{e}tiques
Intenses, CNRS-UGA-UPS-INSA-EMFL, 25 Av. des Martyrs, 38042
Grenoble, France}

\author{W.~Langbein}
\affiliation{School of Physics and Astronomy, Cardiff University,
The Parade, Cardiff CF24 3AA, United Kingdom}

\author{M.~Potemski}
\affiliation{Laboratoire National des Champs Magn\'{e}tiques
Intenses, CNRS-UGA-UPS-INSA-EMFL, 25 Av. des Martyrs, 38042
Grenoble, France} \affiliation{Faculty of Physics, University of
Warsaw, ul. Pasteura 5, 02-093 Warszawa, Poland}

\author{J.~Kasprzak}
\email[]{jacek.kasprzak@neel.cnrs.fr}
\affiliation{Univ. Grenoble Alpes, F-38000 Grenoble, France}
\affiliation{CNRS, Institut
N\'{e}el, "Nanophysique et semiconducteurs" group, F-38000 Grenoble,
France}


\begin{abstract}

Scientific curiosity to uncover original optical properties and
functionalities of atomically thin semiconductors, stemming from
unusual Coulomb interactions in the two-dimensional geometry and
multi-valley band structure, drives the research on monolayers of
transition metal dichalcogenides (TMDs). While recent works
ascertained the exotic energetic schemes of exciton complexes in
TMDs, we here employ four-wave mixing microscopy to indicate that
their subpicosecond dynamics is determined by the surrounding
disorder. Focusing on a monolayer WS$_2$, we observe that exciton
coherence is lost primarily due to interaction with phonons and
relaxation processes towards optically dark excitonic states.
Notably, when temperature is low and disorder weak excitons large
coherence volume results in huge oscillator strength, allowing to
reach the regime of radiatively limited dephasing and we observe
long valley coherence. We thus elucidate the crucial role of exciton
environment in the TMDs on its dynamics and show that revealed
mechanisms are ubiquitous within that family.

\end{abstract}


\date{\today}

\maketitle


\paragraph{\textbf{Introduction}}

In spite of their illusory academic simplicity, synthetic
two-dimensional (2D) materials - such as graphene, black
phosphorous, and transition metal dichalcogenides (TMDs) - display
stunning properties, which are also revealed in their optical
responses. For instance, in monolayers (MLs) of TMDs, the reduced dielectric
screening and 2D carrier confinement give rise to exotic,
non-hydrogenic excitons with binding energies exceeding
0.2\,eV\,\cite{ChernikovPRL14}, which is an asset enhancing
light-matter interaction. The latter is manifested by a strong
absorption and subpicosecond population lifetime, favoring formation
of surface plasmon polaritons\,\cite{Zhou17} and
exciton-polaritons\,\cite{LiuNatPhot14,DufferwielNatCom15,LundtNatCom15}
with a valley degree of freedom - to name a few examples
illustrating a technology-driven progress in the optics of TMDs.
However, there is a need for an in-depth understanding of
fundamental mechanisms governing exciton radiative and nonradiative
recombination rates in various experimental settings. There is a
large spread of reported values of exciton coherence and population decay\,\cite{MoodyJOSAB16} and little is known about
their dependence on microscopic material properties and
environmental factors, such as temperature, strain, dielectric
surrounding and excitonic disorder on different length scales,
generating inhomogeneous broadening $\sigma$.

The main obstacle to access these information, was a large size of
the optically probed areas (typically, diameter of a few tens of
micron), which are required to implement traditional approaches of
nonlinear spectroscopy - such as angle-resolved four-wave mixing
(FWM) - inferring decay times of populations $T_1$ and coherent
polarizations $T_2$ in extended samples. We here overcome this
difficulty, by exploiting phase-sensitive heterodyne detection. The
latter permits to perform FWM spectroscopy in a microscopy
configuration, attaining spatial resolution of $300\,$nm. Using a
tungsten disulphide (WS$_2$) ML, exhibiting the strongest optical
activity among all other TMDs ML\,\cite{LiPRB14} we observe a giant
FWM response of the resonantly generated excitons and we carry out
the mapping of their dephasing time (T$_2=2\hbar/\gamma$),
population decay time and $\sigma$. We further infer the dephasing
induced by phonons, by performing FWM temperature dependence.

Additionally, two distinct types of trions (bound states of two
electrons and one hole) are unambiguously identified in FWM. We show
that a single electron is the ground state for optically active
trions, where the additional electron and hole are within the same
valley as this ground state electron (intra-valley trion), or in the
opposite valley (inter-valley trion). An energetic splitting between
these states due to exchange interaction was recently
predicted\,\cite{YuNatCom2014}, and observed\,\cite{JonesNatPhys16}
for WSe$_2$ and for WS$_2$\,\cite{PlechingerNatComm16}. We observe
the Raman quantum beats\,\cite{FerrioPRL98} resulting from this
splitting, revealing coupling between both types of trions. We
employ this phenomenon to measure the decay of the valley coherence
$T_2^{\text{valley}}$\,\cite{JonesNatNano13,HaoNatPhys16,Hao2DMat17,YeNatPhys17},
which appears to be signiicantly longer that previously reported.

\paragraph{\textbf{Spectral characteristics of transitions}}

\begin{figure}[t]
\includegraphics[width=1.03\columnwidth]{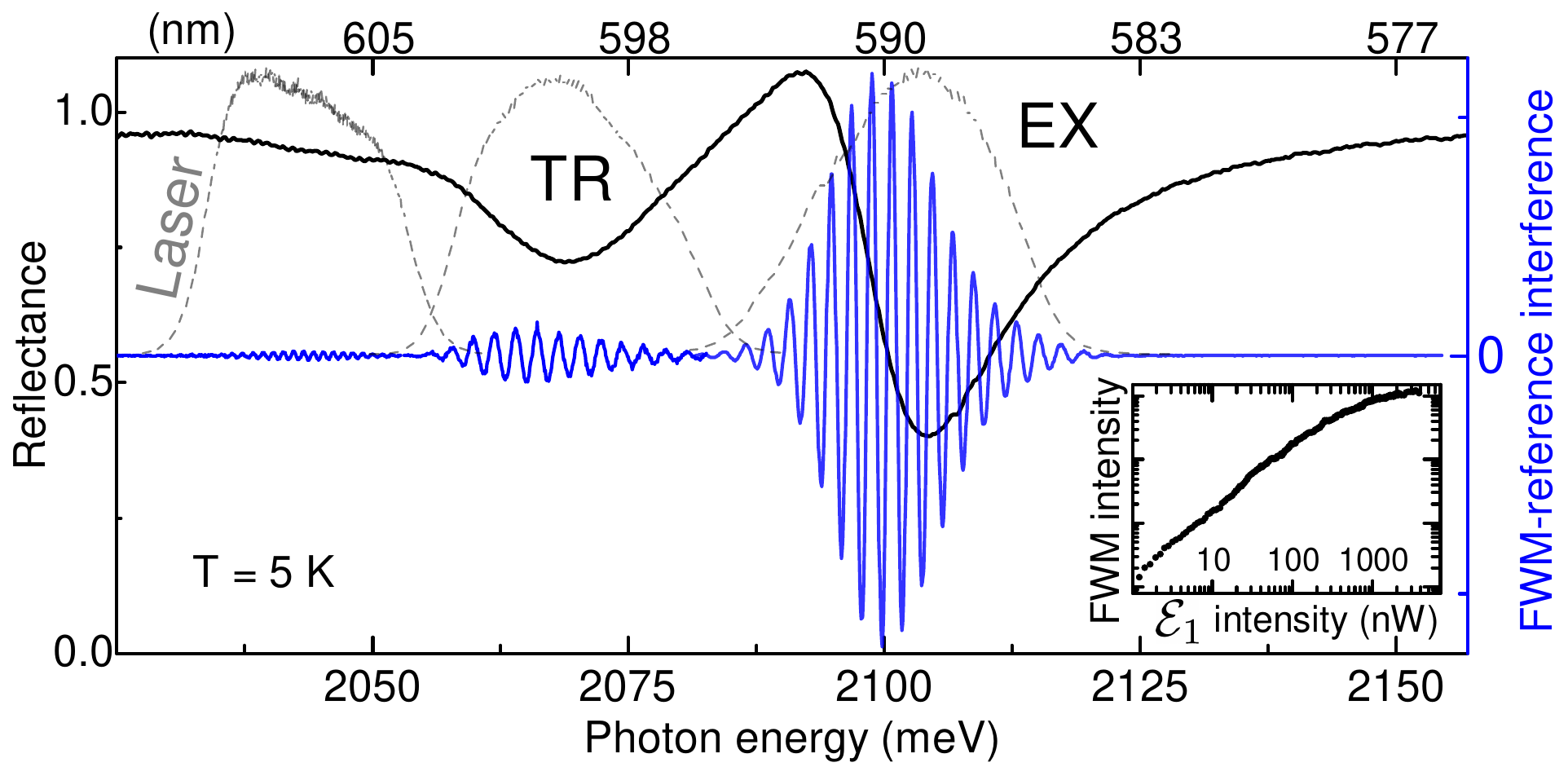}
\caption{{\bf Four-wave mixing micro-spectroscopy of a WS$_2$
monolayer.} Spectrally-resolved reflectance and FWM-reference
interference at $\tau_{12}=0.2\,$ps centered at exciton (EX), trion
(TR) transitions and the third (trionic) transition. The
corresponding spectra of the excitation laser are given by gray
dashed lines. Inset: FWM intensity dependence of EX.
\label{fig:Fig_Spectral_properties}}
\end{figure}

We first perform micro-reflectance from a flake, to identify exciton
(EX) and trion (TR) transitions, as shown in
Fig.\,\ref{fig:Fig_Spectral_properties}. In further experiments we
employ the FWM micro-spectroscopy setup\,\cite{LangbeinOL06,
FrasNatPhot16, JakubczykNanoLett16} (see \itshape{Methods}\upshape),
adapted to the visible spectral range. We probe the flake with the
$\Eo$ pulsed laser beams which are spectrally centered at either EX
($\sim590\,$nm) or TR ($\sim600\,$nm), with a bandwidth about 7\,nm
(full width at half maximum, FWHM). The reference $\Er$ beam is
focused on the surrounding SiO$_2$, so that its lineshape is not
affected by a strong absorption of the flake.
Fig.\,\ref{fig:Fig_Spectral_properties} presents the resulting FWM
spectral interferograms obtained on both resonances in the WS$_2$
flake at T=5\,K. We note that the amplitude of the TR is typically
an order of magnitude weaker than the EX's one. Below the TR line we
further retrieve FWM of another type of
valley-trion\,\cite{MolasNanoscale17}. The fringe period in
Fig.\,\ref{fig:Fig_Spectral_properties} is given by a few
pico-seconds (ps) delay between the reference pulse $\Er$ and the
FWM emission. Its intensity and phase are retrieved by applying
spectral interferometry. The former as a function of $\Ea$ intensity
is shown in the inset, yielding the limit of the third-order
$\chi^{(3)}$ regime (where further experiments are performed) up to
around $100\,$nW. It is worth to note that FWM can be readily
detected with $\Ea$ as low as 1\,nW, generating a low carrier
density of a few $10^8\,$cm$^{-2}$. At excitation stronger than
$100\,$nW, FWM visibly starts to saturate, which we attribute to the
absorption bleaching. In WS$_2$ MLs, the optically active exciton
(EX) has a larger transition energy than the dark
one\,\cite{Molas2DMat17}, such that at low temperature the PL of EX
is suppressed\,\cite{PlechingerNatComm16}, as shown in the
Supplementary Fig.\,S1. While this issue remains relevant in view of
competing relaxation channels of the bright exciton, it is not an
obstacle to drive its FWM: EX are resonantly and selectively
created, generating a giant response, owing to the $\mu^4$ scaling
of the FWM, where $\mu$ is the oscillator strength.

\paragraph{\textbf{Homogenous and inhomogeneous widths of the neutral exciton}}

\begin{figure}[t]
\includegraphics[width=1.03\columnwidth]{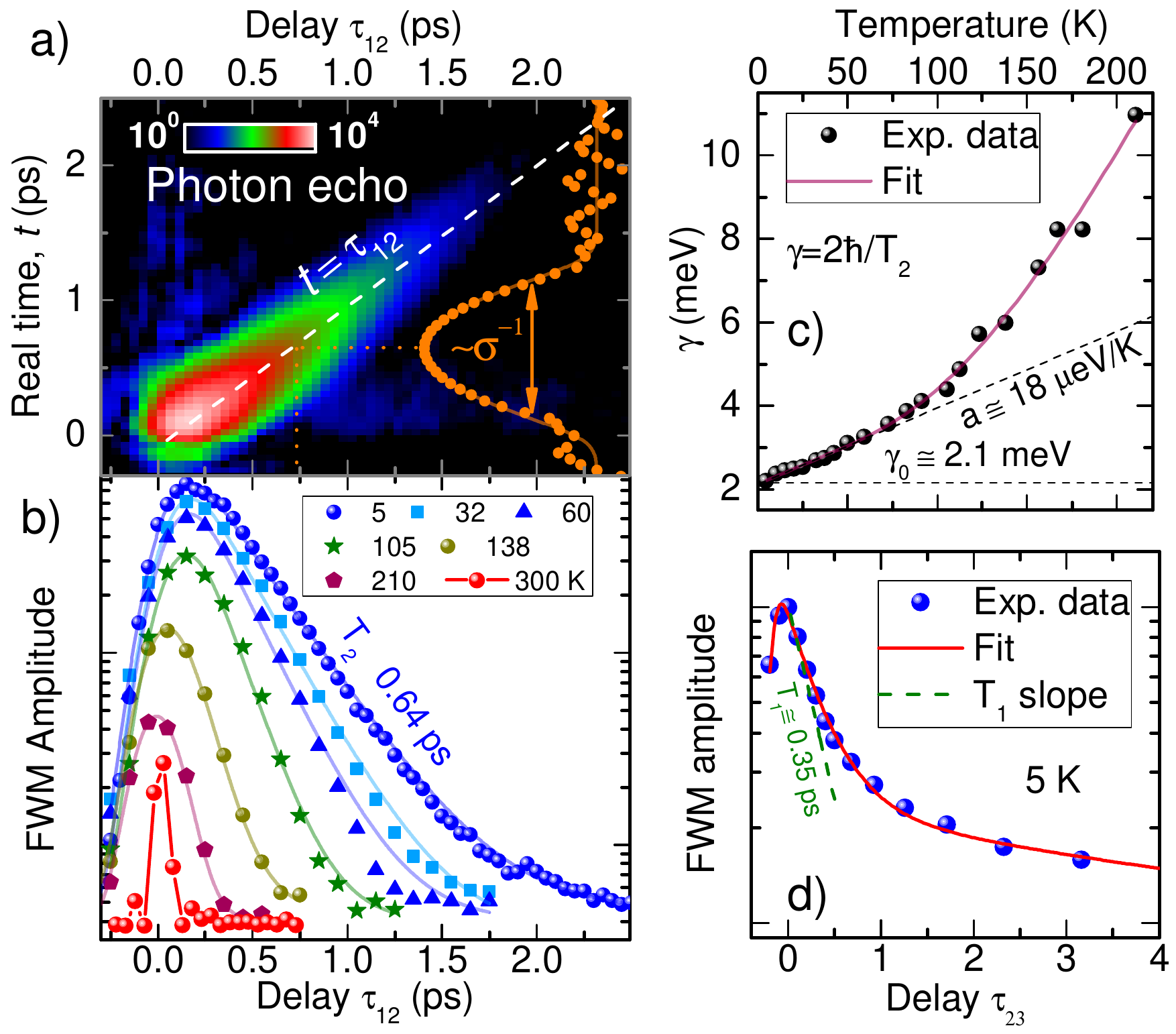}
\caption{{\bf Exciton dephasing in WS$_2$ ML measured in two-beam
FWM.} a)\,Time-resolved FWM showing the photon echo at T$=5\,$K,
$\sigma$ is retrieved from its temporal width (orange trace).
Logarithmic color scale. b)\,Time-integrated echo as a function of
delay $\tau_{12}$ for temperatures, as indicated. The
non-exponential behavior is due to a constant offset, giving a noise
level. c) The retrieved homogeneous broadening versus temperature
$\gamma({\rm T})$ fitted with model described in the text, dashed
lines represent parameters obtained from the fit. d)\,Exemplary
measurement of the EX density dynamics for a given position. The
initial decay T$_{1}$ is mainly due to radiative recombination
T$_{rad}$ and non-radiative relaxation T$_{dark}$ to the EX dark
state, such that (T$_{1})^{-1}$=(T$_{rad})^{-1}$+(T$_{dark})^{-1}$.
\label{fig:Fig_Dephasing_vs_Temp_v1}}
\end{figure}

The EX spectral lineshape measured in reflectance
(Fig.\,\ref{fig:Fig_Spectral_properties}) is dominated by
inhomogeneous broadening with a Gaussian distribution of around
15\,meV FWHM. FWM spectroscopy has been conceived primarily to
access the homogeneous broadening FWHM $\gamma$ in an
inhomogeneously broadened ensemble, exhibiting a spectral standard
deviation of $\sigma$. The complex conjugate in the FWM definition,
imposes phase conjugation between the first-order polarization
induced by $\Ea$ and the FWM. For $\tau_{12}>\hbar/\sigma$, its
transient appears as a Gaussian, known as a photon echo. It is
centered at $t=\tau_{12}$ and has a FWHM, corrected with respect to
the pulse duration, equal to $8\ln{(2)}\hbar/\sigma$. Formation of
such an echo is illustrated in
Fig.\,\ref{fig:Fig_Dephasing_vs_Temp_v1}\,a, where time-resolved FWM
amplitude of EX versus $\tau_{12}$ is shown. The echo develops
during the initial $0<\tau_{12}<0.5$\,ps. By inspecting FWM for
later delays, for example $\tau_{12}=0.7\,$ps (orange trace) we
retrieve inhomogeneous width of around 11\,meV (FWHM).

Time-integrated amplitudes of the photon echo as a function of
$\tau_{12}$ for different temperatures are reported in
Fig.\,\ref{fig:Fig_Dephasing_vs_Temp_v1}\,b. Here, the decay
reflects $\gamma$. To retrieve $\gamma$, the data are fitted with an
exponential decay $\exp{(-\gamma\tau_{12}/2\hbar)}$, convoluted with
a Gaussian to account for a pulse duration of $0.12\,$ps. At T=5\,K
we obtain $\gamma_{\rm{EX}}=(2.1\,\pm\,0.1)\,$meV, yielding
dephasing time T$_2=(620\,\pm\,20)\,$fs. Thus EX in WS$_2$ shows a
larger homogenous width than its counterpart in recently
investigated MoSe$_2$ MLs\,\cite{JakubczykNanoLett16}, in line with
a superior linear absorption in WS$_2$ with respect to
MoSe$_2$\,\cite{LiPRB14}. The temperature dependence of $\gamma$ is
illustrated in Fig.\,\ref{fig:Fig_Dephasing_vs_Temp_v1}\,c. The data
are modeled \,\cite{RudinPRB90}(purple trace) with the following
equation: $\gamma({\rm T})=\gamma_0+a{\rm T}+b/[\exp(E_1/k_{B}{\rm
T})-1]$. The linear term [$\gamma_0=(2.1\,\pm\,0.1)\,$meV,
$a=(0.018\,\pm\,0.003)\,$meV/K] is attributed to low energy acoustic
phonons. The latter term, with the $b=(32\,\pm\,6)\,$meV, is
attributed to thermal activation of optical phonons with dominant or
mean energy $E_1=(37\,\pm\,3)\,$meV, which indeed supply a large
density of states above
$300\,$cm$^{-1}\simeq37\,$meV\,\cite{MolinaPRB11}. The phonon
dephasing mechanisms are therefore similar as in MoSe$_2$
MLs\,\cite{JakubczykNanoLett16} and as in semiconductor quantum
wells. Above T$=210\,$K, the FWM decay is limited by the temporal
resolution, such that $\gamma$ cannot be extracted, although a
strong FWM is measured up to the room temperature.

\paragraph{\textbf{Population decay}}
A representative measurement of the population dynamics (FWM
response versus $\tau_{23}$) via three-beam
FWM\,\cite{FrasNatPhot16,JakubczykNanoLett16} is shown in
Fig.\,\ref{fig:Fig_Dephasing_vs_Temp_v1}\,d. The measurement was
performed at the same spot as the dephasing study, presented in
Fig.\,\ref{fig:Fig_Dephasing_vs_Temp_v1}. The population dynamics is
dominated by an initial exponential decay with a constant of T$_1
\simeq 0.35$\,ps, followed by a longer dynamics described by two
additional exponential decays\,\cite{Scarpelli17}
(T$^1_{\text{slow}} \simeq 4.7$\,ps and T$^2_{\text{slow}} \simeq
46$\,ps) that we can relate to phase space distribution via
scattering processes and scattering back from the exciton dark
ground state.
 We note that the
portion of secondary excitons, decaying  on a nano-second timescale,
is at least an order of magnitude larger than on recently studied
MoSe$_2$ MLs. This we associate with a dark exciton ground state in
WS$_2$ and its bright character in MoSe$_2$\,\cite{Molas2DMat17}.

The obtained result (T$_{2}\simeq 2$T$_{1}$) indicates
population-limited dephasing. We attribute this fast decay to be due
primarily to the fast radiative decay. Indeed, excitons in ML TMDs
possess the radiative lifetime T$_{rad}$ of a few hundred
femto-seconds, as recently revealed via two-colour
pump-probe\,\cite{PoellmannNatMat14}, FWM\,\cite{MoodyNatCom15,
JakubczykNanoLett16} and TMD polaritons
studies\,\cite{LiuNatPhot14,DufferwielNatCom15,LundtNatCom15} - all
these results signify a large EX transition dipole moment and
coherence volume spanning across many Bohr
radii\,\cite{FeldmannPRL87}. The parameter T$_{dark}$ describing
phase space distribution via scattering processes and relaxation to
the dark exciton ground state is also expected to contribute to the
fast initial decay. Other nonradiative recombination processes are
expected to be of minor impact, as they are not faster than the
decay of secondary excitons, that is $\geq$46\,ps (we assume that
these processes have the same dynamics for both bright and dark
excitons). We also note a weak role of phonons on the excitons
dynamics in this low temperature range (see
Fig.\,\ref{fig:Fig_Dephasing_vs_Temp_v1}\,c). To get a comprehensive
view of the possible mechanisms influencing the exciton dynamics,
the local insight into  $\sigma$, T$_2$, population decay and $\mu$
is required, and should be strengthen by imaging of these quantities
across the entire flake. Crucially, such an original capability is
offered by the heterodyne FWM microscopy. Thus, we now focus on the
FWM mappings and analyze spatial correlations between the above
parameters.

\begin{figure}[t]
\includegraphics[width=1.03\columnwidth]{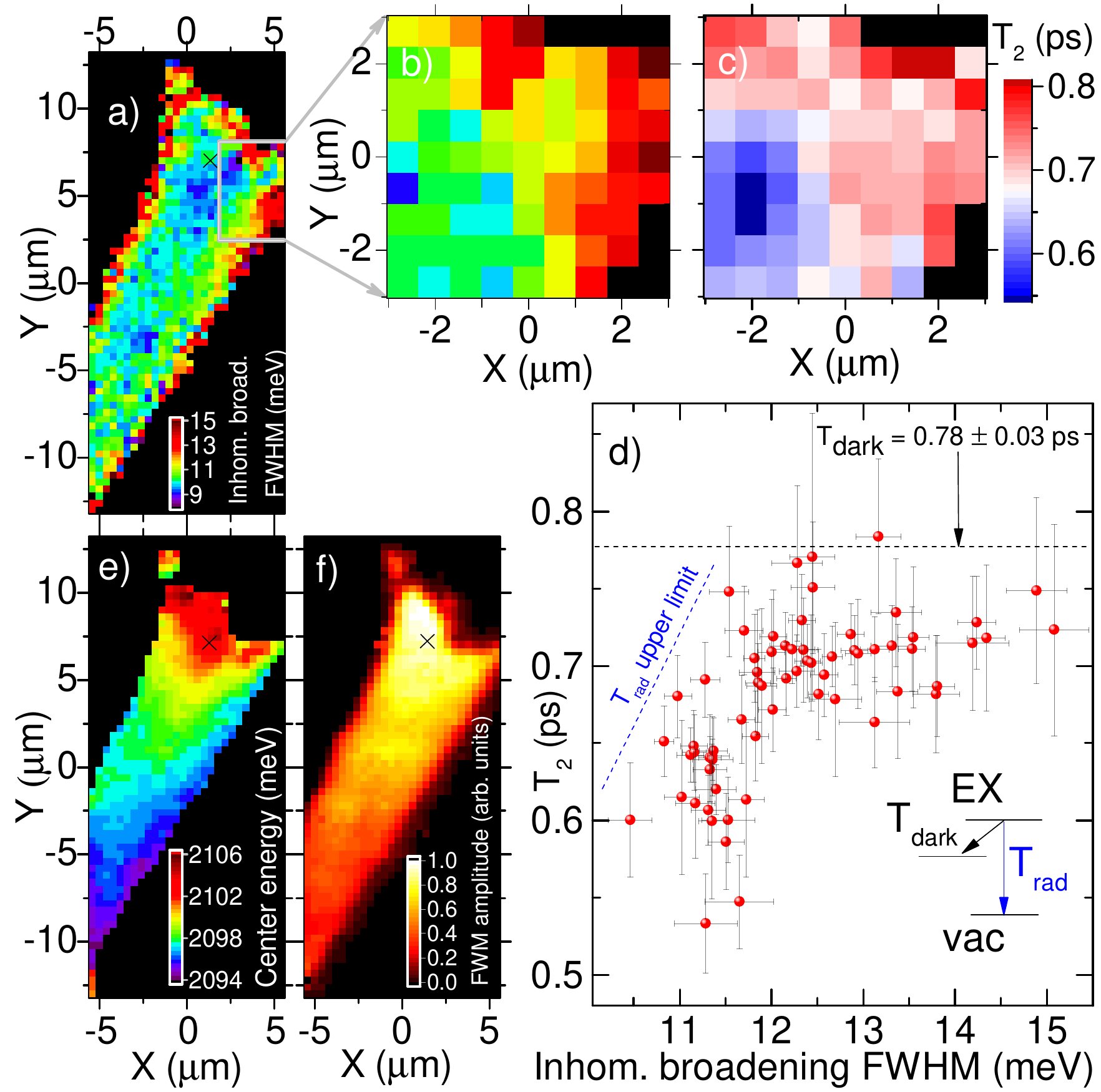}
\caption{{\bf Spatially-resolved probing of EX coherence in a WS$_2$
ML via FWM hyperspectral imaging.} a)\,Spatial imaging of the
inhomogeneous broadening $\sigma$ retrieved from time-resolved FWM
at $\tau_{12}=0.6\,$ps. b)\, Zoomed map of inhomogeneous broadening
and (c) dephasing time. d)\,Correlation between homogeneous and
inhomogeneous broadenings, indicating radiative rate governed by
localization via disorder and illustrating competition between
radiative recombination and non-radiative EX shuffling to its dark
state (sketched). The dashed black line indicates $T_{dark}$
extracted from a fit (see the Supporting Information).
e)\,Hyperspectral FWM imaging, $\tau_{12}=0.2\,$ps, hue indicates
transition energy. f)\,Time-integrated FWM amplitude reflecting
$\mu$, linear scale over an order of magnitude. The black cross
indicates location where results presented on
Figs.\,\ref{fig:Fig_Dephasing_vs_Temp_v1} and \ref{fig:FigTrion}
were obtained. \label{fig:Fig_Mappings}}
\end{figure}

\paragraph{\textbf{Four-wave mixing mapping}}
In the first step, we focus on spatial variation of $\sigma$. We
therefore acquire FWM spectral interferograms at $\tau_{12}=0.6\,$ps
and retrieve time-resolved FWM amplitude of EX, while scanning over
the flake surface. For each location, we inspect the width of the
photon echo, from which we measure $\sigma$. The result is shown in
Fig.\,\ref{fig:Fig_Mappings}\,a. In the middle of the flake, we
identify regions of a smaller inhomogeneous broadening, down to
10\,meV, yet still largely dominating over $\gamma$. It is worth to
note, that the largest $\sigma$, and thus most pronounced exciton
localization, is measured at the borders of the flake. This is
related to the strain gradients and variations of the dielectric
screening by the substrate, which are expected to be strong along
the edges. These locations are preferential for wrinkling, local
deformations and lattice defects creating deep potential centers
trapping individual emitters (see supplementary Fig.\,S1\,d). As an
origin of $\sigma$, we point toward a local strain and charges
trapped on a flake. We note that suspended ML flakes displayed
comparable $\sigma$ (see supplementary Fig.\,S3), excluding the
interface roughness between the SiO$_2$ and the flake as a principal
source of inhomogeneity. Newly, it has been found that $\sigma$ is
reduced by encapsulating a flake in hexagonal boron
nitride\,\cite{Cadiz17, Ajayi17}: FWM performed on such
heterostructure indeed has revealed reduction of $\sigma$, however
its complete cancelation has not been observed (not shown).
Moreover, a reduced spectral jitter was observed on non-insulating
substrates\,\cite{IffOptica17}, helping to get rid of trapped
charges, further indicating decisive role of charge fluctuation on
the inhomogeneous broadening.

In the next step, we focus on the area exhibiting a large variation
of $\sigma$, marked with a square in
Fig.\,\ref{fig:Fig_Mappings}\,a. Again we perform mappings, this
time varying also the delay $\tau_{12}$ for each position. Such
retrieved T$_2$ and $\sigma$ are presented as color-coded maps in
Fig.\,\ref{fig:Fig_Mappings}\,b and c, respectively. Their spatial
correlation is striking and emphasized in
Fig.\,\ref{fig:Fig_Mappings}\,d: the fastest dephasing is measured
at the areas of smallest $\sigma$. We interpret this as follows.
Center of mass of two-dimensional excitons moves within a disordered
potential landscape\,\cite{SavonaPRB06}, arising from the variation
dielectric contrasts, strain from the substrate, uncontrollable
impurities, vacancies, etc. Through the Shr\"{o}dinger equation, the
disorder acts on the wave-function localization in real space and
thus results in its delocalization in k-space, modifying radiative
rates with respect to free excitons. In other words, the disorder
mixes the states inside and outside of the radiative cone, and thus
creates a distribution of states with an oscillator strength reduced
as compared to ones fully in the radiative cone, and spread in
energy, adding up to $\sigma$. Note, that this localization is weak
\cite{SinghPRB16} comparing to localization resulting from deep
traps \cite{HichriJAP17}, resulting in a distinctive emission band
well below the EX emission (see Supplementary Information
Fig.\,S1\,d). As presented in Fig.\,\ref{fig:Fig_Mappings}\,d, T$_2$
starts to decrease only for a sufficiently low $\sigma$, where the
radiative decay time T$_{rad}$ becomes fast enough to compete with
another channel, identified as the EX relaxation to the dark ground
state. Such channel was not observed in MoSe$_2$ displaying a bright
exciton ground state (see Supplementary Fig.\,S3\,a) and MoS$_2$
(not shown). Conversely, for largest $\sigma$, the non-radiative
decay dominates, as the T$_{rad}$ is increased through the
localization. These spatial correlations demonstrate that radiative
rates and dephasing of excitons in ML of TMDs are governed by
exciton localization imposed by a local disorder.

The slope of the transition energy owing to the strain gradient is
observed in Fig.\,\ref{fig:Fig_Mappings}\,e,  where the center
transition energy is encoded in a hue level. In
Fig.\,\ref{fig:Fig_Mappings}\,f we present time-integrated FWM
amplitude of the EX transition (corrected by the excitation
lineshape) reflecting $\mu$. Comparing
Fig.\,\ref{fig:Fig_Mappings}\,f with
Fig.\,\ref{fig:Fig_Mappings}\,a, we note that the areas of the
smallest $\sigma$ yield the strongest FWM (see also Supplementary
Fig.\,3). This is because with decreasing $\sigma$ (disorder), the
spatial overlap between excitons increases, enhancing the EX
interaction strength and thus resulting in a more intense
time-integrated FWM. We note that the largest oscillator strength is
observed in areas with the lowest transition energy.

\paragraph{\textbf{Valley-trion structure}}

\begin{figure}[t]
\includegraphics[width=1.03\columnwidth]{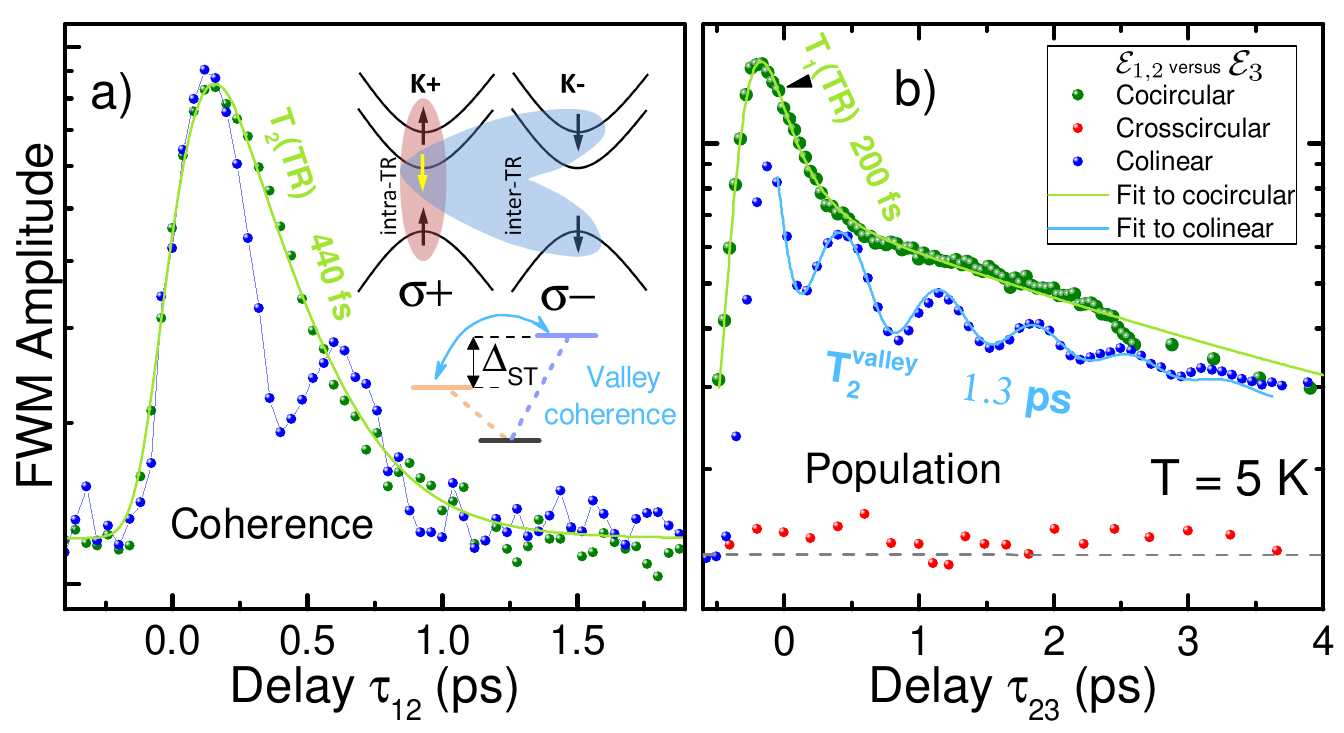}
\caption{{\bf FWM of the intra- and inter-valley trions in a WS$_2$
ML.} a)\,Trion coherence dynamics measured in co-circular (brown)
and co-linear (green) polarizations of $\Eo$. Inset: Intra-TR in K+
valley and inter-TR in K- valley are Raman-coupled by sharing the
same ground state (yellow electron, symbolized with a down-arrow ).
b)\,Population dynamics measured in co-circular and co-linear
polarizations of $\Eo$. Semi-transparent traces are the fitted
dynamics. Owing to $\sigma$, the measured contrast of oscillations
is less than simulated. For the cross-circular setting of $\Ed$ and
$\Ec$ the measured FWM is at the noise level (gray dashed line),
indicating a strongly suppressed inter-valley scattering.
\label{fig:FigTrion}}
\end{figure}

We now turn to investigation of the trion transition
(TR)\,\cite{PlechingerNatComm16}. The latter is formed when an
additional electron occupies the lowest conduction band, as depicted
in the inset of Fig.\,\ref{fig:FigTrion}\,a. Depending on its spin,
one can form a singlet state (intra-valley trion, intra-TR) or a
triplet-state (inter-valley trion,
inter-TR)\,\cite{PlechingerNatComm16, SinghPRL16}. To address
coherence dynamics of these TR, $\Eo$ are prepared co-circularly,
selectively addressing K+ valleys. We investigate it at the same spatial location as for experiments illustrated on Fig.\,\ref{fig:Fig_Dephasing_vs_Temp_v1}, yielding low $\sigma$ and marked with a cross on Fig.\,\ref{fig:FigTrion}\,a,\,e and f. Similarly as for EX, we obtain a
single exponential decay yielding the averaged TR dephasing T$_2(\rm
TR)=(440\,\pm\,10)\,$fs. This faster dephasing with respect to EX is
attributed to fast TR relaxation into the lower lying dark states,
leaving an electron with a varying momentum, and inducing additional
dephasing via final state damping.

When $\Eo$ are co-linear, K+ and K- valleys are equally excited, as
linearly polarized light contains both circularly polarized
components. As a result, intra-TR and inter-TR transitions are
activated in both valleys. In particular, intra-TR in K+ and
inter-TR in K- valley share the same ground state (corresponding to
the presence of an electron in the lowest conduction band, labeled
with a yellow arrow in the inset in Fig.\,\ref{fig:FigTrion}\,a),
forming a coupled V-type system. In such a configuration, $\Ea$ and
$\Eb$ generate valley coherence between both types of trions
resulting in the Raman quantum beats\,\cite{FerrioPRL98} , as
sketched in the inset in Fig.\,\ref{fig:FigTrion}\,a. Owing to the
TR singlet-triplet splitting\,\cite{PlechingerNatComm16}, labeled as
$\Delta_{\rm ST}$, the phase of this coherence evolves when
increasing $\tau_{12}$. Therefore, the measured coherence dynamics,
shown in Fig.\,\ref{fig:FigTrion}\,a, displays
beating\,\cite{FerrioPRL98} with a period $T_{\rm
ST}\simeq0.71\,$ps, yielding $\Delta_{\rm ST}=2\pi\hbar/T_{\rm
ST}=5.8\,$meV.

To measure the TR density dynamics we perform the
$\tau_{23}$-dependence of the FWM\,\cite{JakubczykNanoLett16} for a
fixed $\tau_{12}=0.2\,$ps. Once again, via co-circular excitation we
selectively probe the intra-TR dynamics. As presented in
Fig.\,\ref{fig:FigTrion}\,b, it displays a fast monotonous decay,
owing to the relaxation into the dark exciton state, followed by a
slower revival of the FWM generated by the dark excitons relaxing
back into the light cone\,\cite{JakubczykNanoLett16}. This
interpretation is strengthened by comparing the TR dynamics in
WS$_2$ with the one measured in MoSe$_2$ ML (see Supplementary
Fig.\,S4), exhibiting optically bright ground state. In the latter
case, as the TR recombination leaves an electron with a varying
momentum and relaxation to lower energy state is not possible, we
observe a substantially longer TR radiative decay with respect to
EX.

It is worth pointing out that no FWM is observed (either at TR or
EX) for opposite circular polarizations of $\Ed$ and $\Ec$ pointing
towards a robust valley polarization in WS$_2$ MLs.

Upon co-linear excitation, the initial density dynamics displays
again an oscillatory behavior: the first two pulses do not only
create the intra-TR and inter-TR populations, but also induce the
valley coherence between them, once more generating the Raman
quantum beats\,\cite{FerrioPRL98}. All these ingredients contribute
to the FWM signals, which we model with a phenomenological fitting
curve (see Supporting Information) and extract the valley coherence
dephasing time of $T_2^{\text{valley}}=(1.3\,\pm\,0.2)\,$ps. Such a
record value, much longer than previously
reported\,\cite{JonesNatNano13,HaoNatPhys16,Hao2DMat17,YeNatPhys17}
and measured on location with low $\sigma$ confirms recent reports
suggesting that a shallow disorder potential plays a critical role
in the exciton valley coherence\,\cite{TranPRB2017}. For longer
delays the valley coherence has dephased, such that the subsequent
exciton dynamics is similar for both polarization configuration.
Thus both TR transitions of WS$_2$ are here unveiled via FWM,
whereas they cannot be distinguished in reflectance owing to the
spectral splitting smeared out by $\sigma_{\rm TR}$. Similar values
of $\Delta_{\rm ST}$ were retrieved when varying the position on the
flake.

To conclude, we demonstrated a giant nonlinear optical response of
exciton complexes in WS$_2$ MLs. The substantial enhancement of the
FWM retrieval efficiency with respect to standard semiconductor
quantum wells, was exploited to unravel the impact of a local
disorder in two-dimensional systems onto exciton dynamics and
dephasing. The valley degree of freedom was unveiled, when
considering the trion structure. Our results indicate that coherent
nonlinear microscopy is suited to explore optical properties of
emerging optoelectronic and optomechanical\,\cite{MorellNL16}
devices and heterostructures\,\cite{Cadiz17} made of layered
semiconductors. An alluring perspective is to image the exciton
coherent dynamics on a nanometer areas and to conjugate it with the
structural properties of TMDs, which can be revealed down to atomic
scale using scanning tunneling microscopy. This could be achieved by
transferring our methodology towards the nanoscopic
regime\,\cite{AeschlimannScinece11, KravtsovNatNano15}, offering a
spatial resolution of a few tens of nanometers.

\section*{Acknowledgement} We acknowledge the financial support by the
European Research Council (ERC) Starting Grant PICSEN (grant no.
306387), the ERC Advanced Grant MOMB (grant no.\,320590), the EC
Graphene Flagship project (No. 604391) and the ATOMOPTO project
within the TEAM programme of the Foundation for Polish Science
co-financed by the EU within the ERDFund. We also acknowledge the
technical support from Nanofab facility of the Institute N\'{e}el,
CNRS UGA.

\section*{Methods}

We employ an optical parametric oscillator (Inspire 50 by Radiantis
pumped by Tsunami Femto by Spectra-Physics) to create a triplet of
short laser pulses around 600\,nm: $\Ea$, $\Eb$ and $\Ec$, with
adjustable delays $\tau_{12}$ and $\tau_{23}$, as depicted in the
Supplementary Fig.\,S1. The three beams are injected co-linearly
into the microscope objective (Olympus VIS, NA=0.6), installed on a
XYZ piezo stage. They are focused down to the diffraction limit of
0.6$\,\mu$m, onto the sample placed in a helium-flow cryostat. $\Eo$
are pre-chirped by using a geometrical pulse
shaping\,\cite{FrasNatPhot16}, so as to attain close to
Fourier-limited, 120\,fs pulses on the sample. The WS$_2$ ML flake
was mechanically exfoliated from a bulk crystal purchased from
HQ-graphene and deposited on a 90\,nm thick SiO$_2$ substrate.

The FWM generated within the sub-wavelength (approximately half of
the waist) area, diffracts in all directions. There is therefore no
k-vector matching condition, on which most FWM experiments rely on.
Instead, our microscopy approach imposes the signal to be selected
in phase, by performing optical heterodyning. By employing
acousto-optic deflectors operating at different radio-frequencies
$\Omega_{1,\,2,\,3}$, the phases within the pulse trains $\Eo$ are
modulated by $n\Omega_{1,\,2,\,3}/\nu$, where $\nu$ and $n$ denote
the laser repetition rate and pulse index within the train,
respectively. As a result, the FWM polarization - which in the
lowest, third-order is proportional to $\mu^4\Ea^{\star}\Eb\Ec$ -
evolves with the phase $n(\Omega_3+\Omega_2-\Omega_1)/\nu$. This
specific phase-drift is locked onto the reference pulse $\Er$,
overlayed with the reflected light, and thus producing a stationary
interference with the FWM field. The background free
interference\,\cite{LangbeinOL06} is spectrally dispersed by an
imaging spectrometer and detected on a CCD camera.


\newpage

\widetext

\begin{center}
{\bf \large SUPPLEMENTARY MATERIAL\\ Impact of environment on
dynamics of exciton complexes in a WS$_2$ monolayer\\} T. Jakubczyk,
K. Nogajewski, M. R. Molas, M. Bartos, W. Langbein, M. Potemski, and
J. Kasprzak
\end{center}

\setcounter{figure}{0}

\renewcommand{\figurename}{Supplementary Figure}
\renewcommand{\thefigure}{S\arabic{figure}}

\section{Rationale of the experiment}
\begin{figure}[!htb]
\includegraphics[width=0.7\columnwidth]{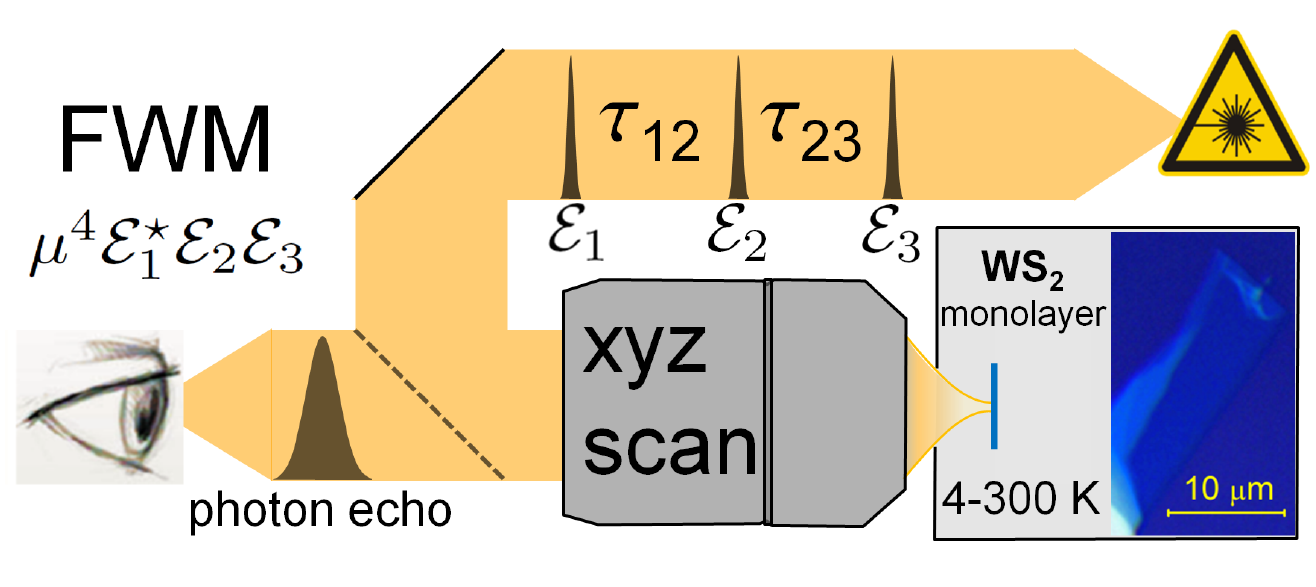}
\caption{\bf Cartoon illustrating rationale of the experiment. The
upper-right pictograph represent the used laser chain (CW 532$\,$nm
$\mapsto$ Ti:Sapphire, femto-second 810\,nm $\mapsto$ OPO,
femto-second pulses in VIS range). The lower-left pictograph (an
eye) symbolizes the detection via heterodyne spectral interferometry
approach\,\cite{LangbeinOL06}.\label{fig:FigS0}}
\end{figure}

\section{Photoluminescence hyperspectral imaging}
\begin{figure}[!htb]
\includegraphics[width=0.53\columnwidth]{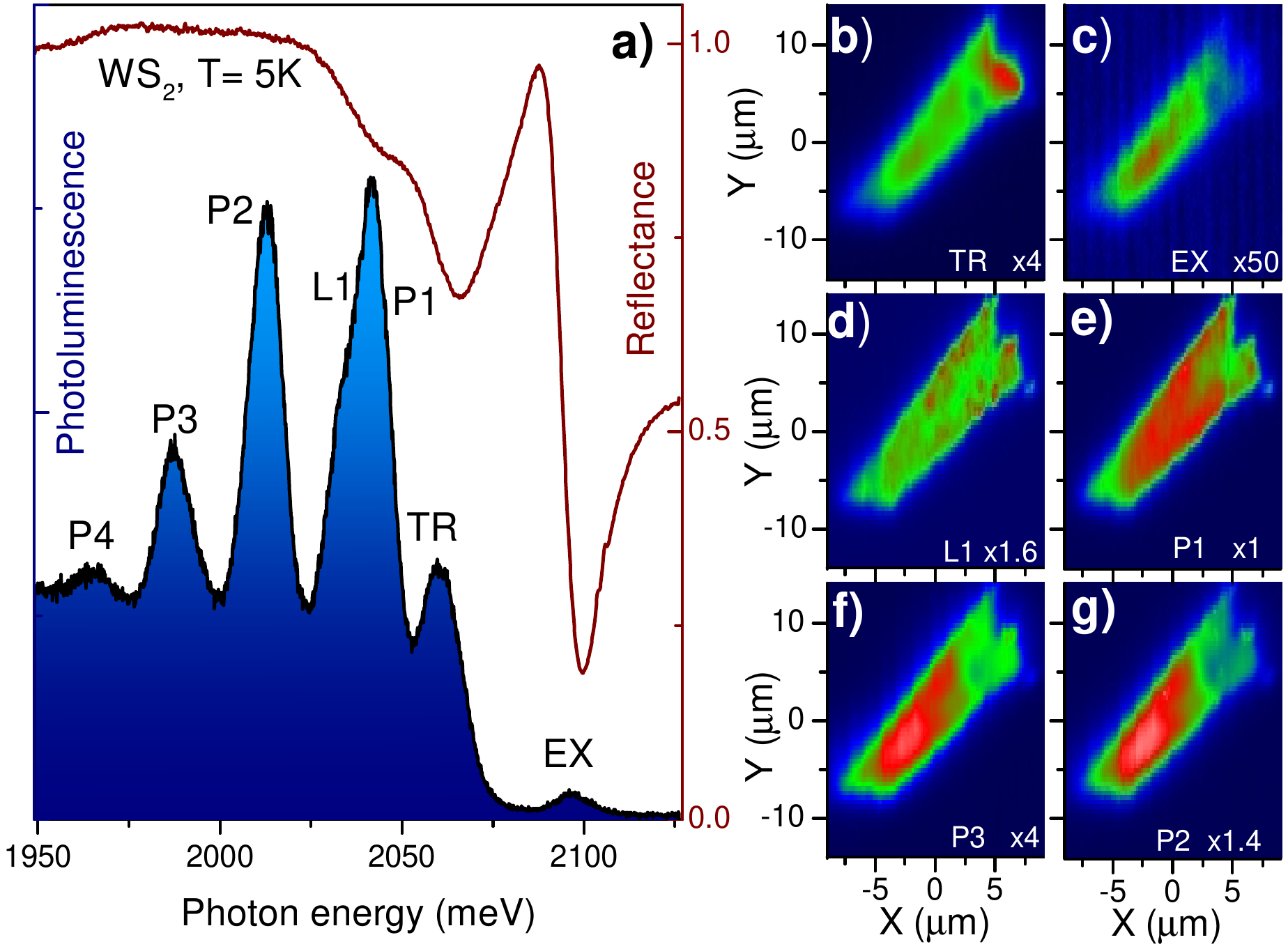}
\caption{{\bf Photoluminescence hyperspectral imaging of a WS$_2$
monolayer at T$=5\,$K.} a)\,As the optically active exciton (EX) is
an excited state, its PL is strongly quenched. TR labels the
negative trion. P1 was recently interpreted as another type of
valley-trion\,\cite{MolasNanoscale17} coinciding with the dark EX.
The origin of peaks P2, P3 and P4 remains unexplained. The
reflectance is given in red, for comparison with the PL. The spatial
imaging spectrally integrated across the peaks, as indicated, is
shown in (b-g). L1 indicates spectrally narrow, strongly localized
states, which are formed at the low energy side of P1. CW excitation
at 450\,nm with $\simeq5\,\mu$W at the sample.\label{fig:FigS1}}
\end{figure}

\section{Complementary results obtained on M\MakeLowercase{o}S\MakeLowercase{e}$_2$ monolayers}
\begin{figure}[!htb]
\includegraphics[width=0.78\columnwidth]{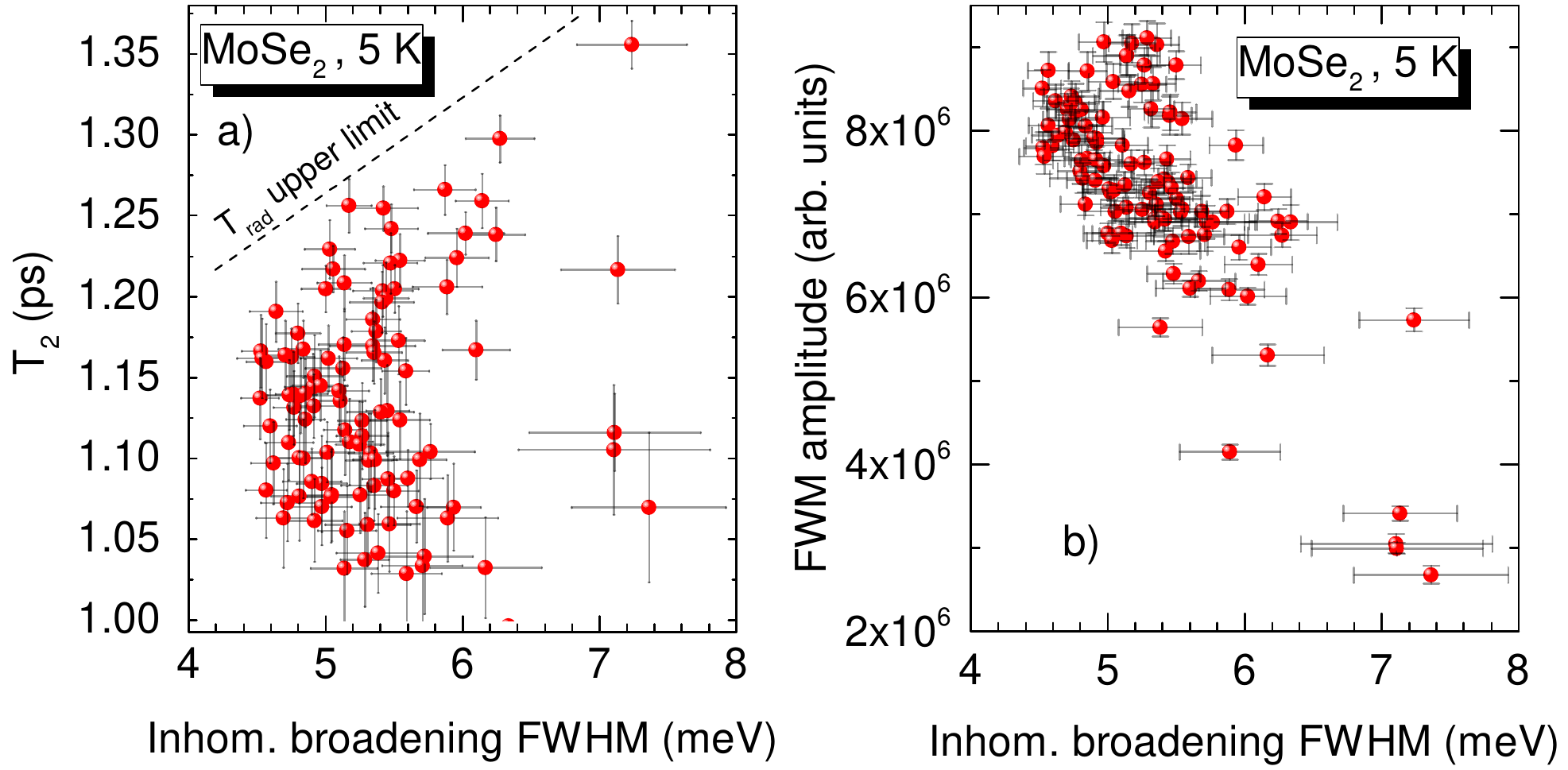}
\caption{{\bf Measurements  at MoSe$_2$ monolayer presented in
Ref.\,[\onlinecite{JakubczykNanoLett16}]} a)\,Correlation between
homogeneous and inhomogeneous broadenings, indicating radiative rate
governed by localization via disorder. Note that comparing to
WS$_{2}$ no saturation is observed for large inhomogeneous
broadening values. We interpret the vertical spread of experimental
values as resulting from activation of non-radiative recombination
and fast scattering out of radiative cone, resulting in faster
decoherence. Therefore, for each value of inhomogeneous broadening,
only the upper point correspond to radiatively limited dephasing.
b)\,Time-integrated FWM amplitude versus inhomogeneous broadening,
the areas of the smallest $\sigma$ yield the strongest FWM. This is
because with decreasing $\sigma$ (disorder), the spatial overlap
between excitons increases, enhancing the EX interaction strength
and thus resulting in a more intense time-integrated FWM response.
\label{fig:MoSe2}}
\end{figure}

\begin{figure}[!htb]
\includegraphics[width=0.45\columnwidth]{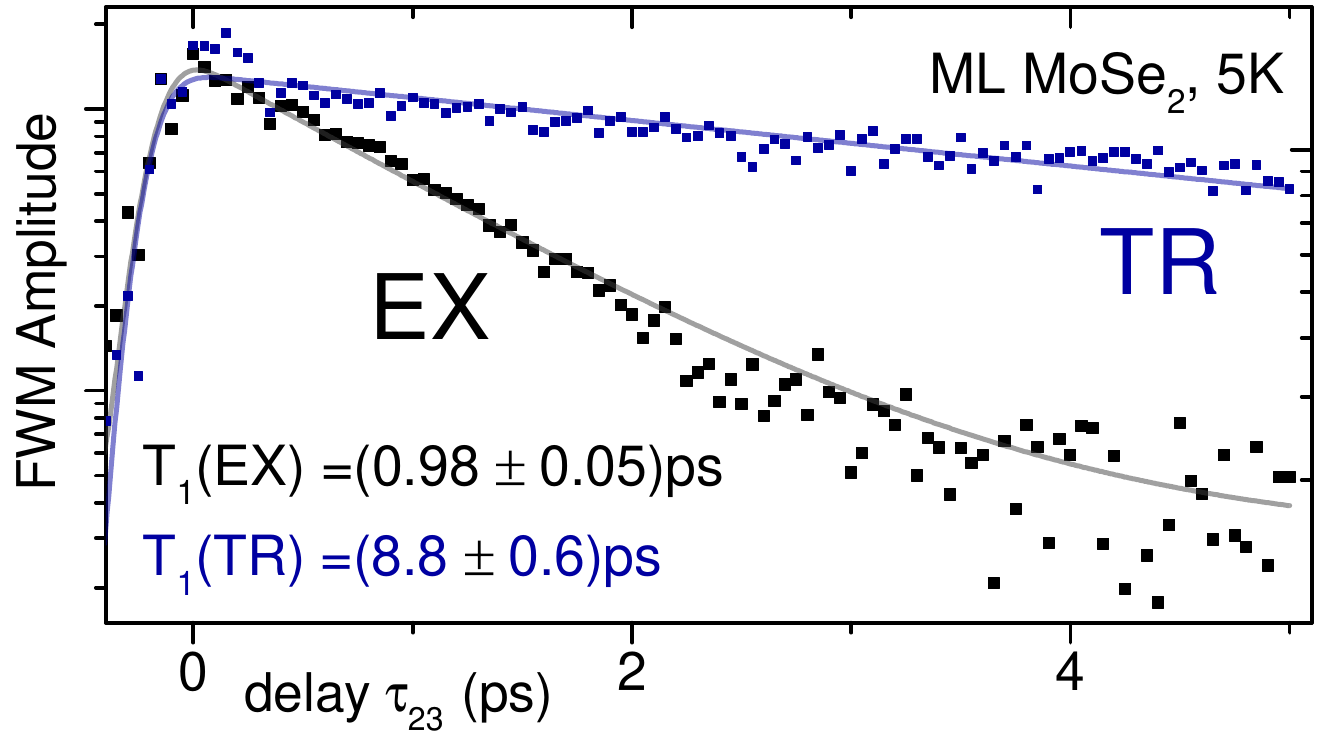}
\caption{{\bf Density dynamics measured at EX and TR in a MoSe$_2$
monolayer, hosting optically bright exciton in the ground state.}
The TR lifetime is an order of magnitude longer than the EX
one.\label{fig:FigS2}}
\end{figure}

\begin{figure}[!htb]
\includegraphics[width=0.6\columnwidth]{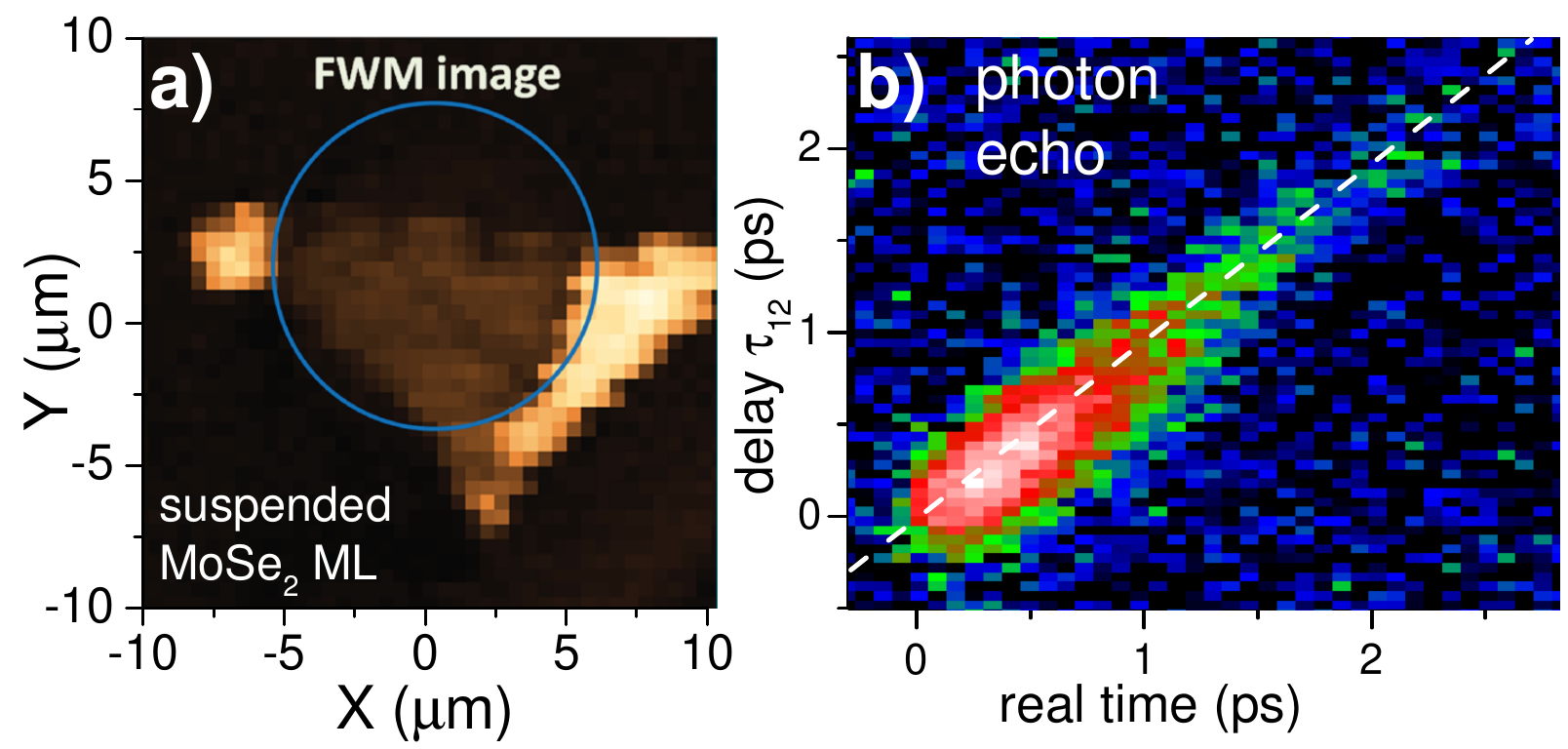}
\caption{{\bf FWM micro-spectroscopy of a suspended MoSe$_2$
monolayer flake.} a)\, Time-integrated FWM imaging of the flake
exfoliated onto a hole-aperture (indicated with a blue circle),
surrounded by a SiO$_2$ substrate. A weakened FWM response is due to
a lower laser field intensity at the aperture with respect to the
substrate. b)\,An exemplary photon echo, measured via two-beam FWM,
at the suspended part of the flake. Similar traces were obtained on
other locations, both at the suspended and non-suspended parts. This
indicated that the contact with the substrate is not the culprit of
the inhomogeneous broadening.\label{fig:FigS3}}
\end{figure}

\section{Mapping of the exciton population dynamics in WS$_2$ employing spatially-resolved three-beam FWM}
\begin{figure}[!htb]
\includegraphics[width=0.4\columnwidth]{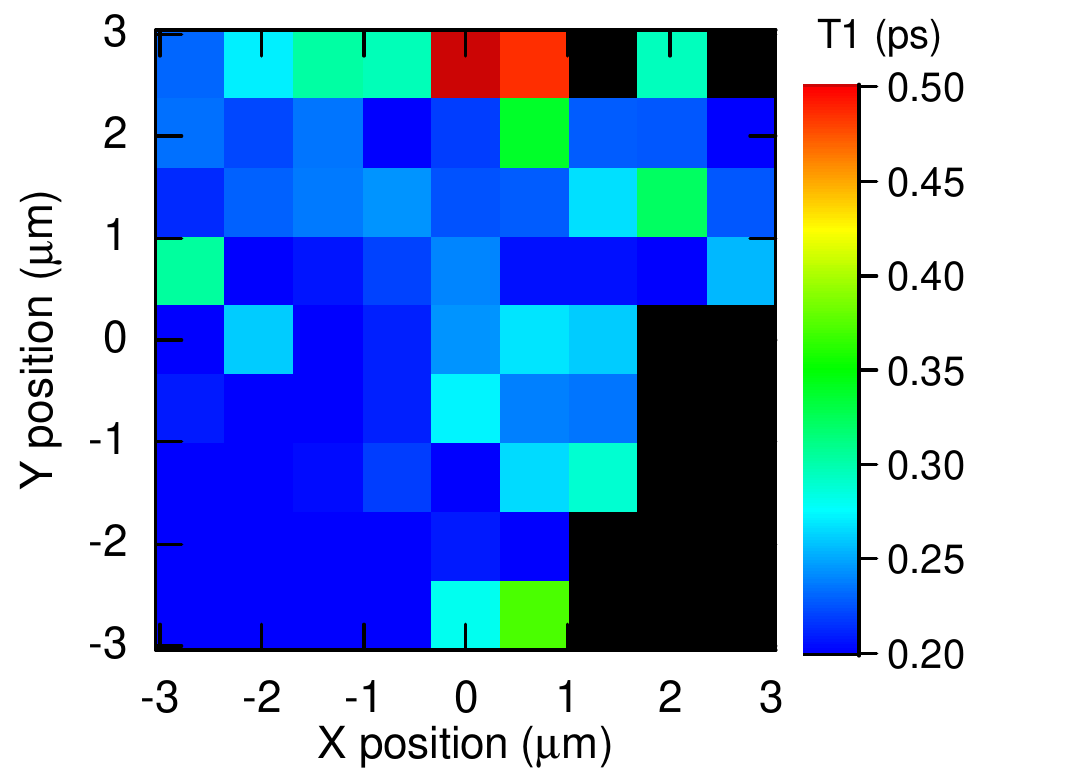}
\caption{\bf Mapping of T$_1$ (the fastest decay component)
performed on the same location as maps on Fig.\,3\,b and c.
Crucially, the areas characterized by a fastest T$_{\rm 1}$ also
display short T$_2$ and the smallest $\sigma$ (compare to
Fig.\,3\,a, b and c). } \label{fig:Fig_T1}
\end{figure}

\section{Extracting of T$_{dark}$ from the T$_{2}$ vs. inhomogeneous broadening dependence}
To extract T$_{dark}$ we fit the following phenomenological
dependence to data presented on Fig.\,3\,d:
\begin{equation}
1/T_2 = (A\sigma+B) + \frac{1}{T_{dark}} \label{eq:phenomenological}
\end{equation}
and obtain T$_{dark}  = 0.78 \pm 0.03$\,ps.

\section{Raman coherence fit}
We fit the FWM amplitude vs $\tau_{23}$ dependence measured in
colinear configuration on the TR, which is presented on Fig.\,4\,b,
with a phenomenological model of the following form:
\begin{equation}
\text{y0} + \left( A e^{-\frac{\tau
_{23}}{\text{T}_{\text{fast}}}}+B e^{-\frac{\tau
_{23}}{\text{T}_{\text{slow}}}} \right) \left(e^{-\frac{\tau
_{23}}{T_2^{\text{valley}}}} \cos \left(\frac{\Delta_{ST}}{\hbar
}\tau _{23}+\phi \right)+2 \right), \label{Eq:Raman_fit}
\end{equation}
where $T_2^{\text{valley}}$ is the valley coherence decay constant,
${\text{T}_{\text{fast}}}$ and ${\text{T}_{\text{slow}}}$ describe
the fast and slow population decay components, respectively, while
$A$ and $B$ are their amplitudes, y$_0$ is the background signal and
$\Delta_{ST}$ is the TR singlet-triplet splitting.
Table\,\ref{my-label} presents the obtained fitting parameters. We
performed fits on data obtained on different spots and values
presented without experimental error here were typical values
obtained from those fits. Therefore, only three free fitting
parameters were used to fit data presented in the article ($A$, $B$
and $T_2^{\text{valley}}$).

\begin{table}[!htb]
\centering \caption{Raman coherence fit results} \label{my-label}
\begin{tabular}{ | l | l | l | l | l | l | l | l | l |  }
\hline
        A & B &  $\Delta_{ST}$ [meV] & $T_{fast}$ [ps] & $T_{slow}$ [ps] & $T_2^{\text{valley}}$ [ps] & $y_0$   \\ \hline
     $6547 \pm 313$ &
     $2247  \pm 15$ &
     $5.76$ &
     $0.2 $ &
     $4 $ &
     $1.33 \pm 0.07$ &
     $1220 $  \\    \hline

\end{tabular}
\end{table}

\section{Trion transition - Feynman diagram of possible quantum pathways}
\begin{figure}[!htb]
\includegraphics[width=0.75\textwidth]{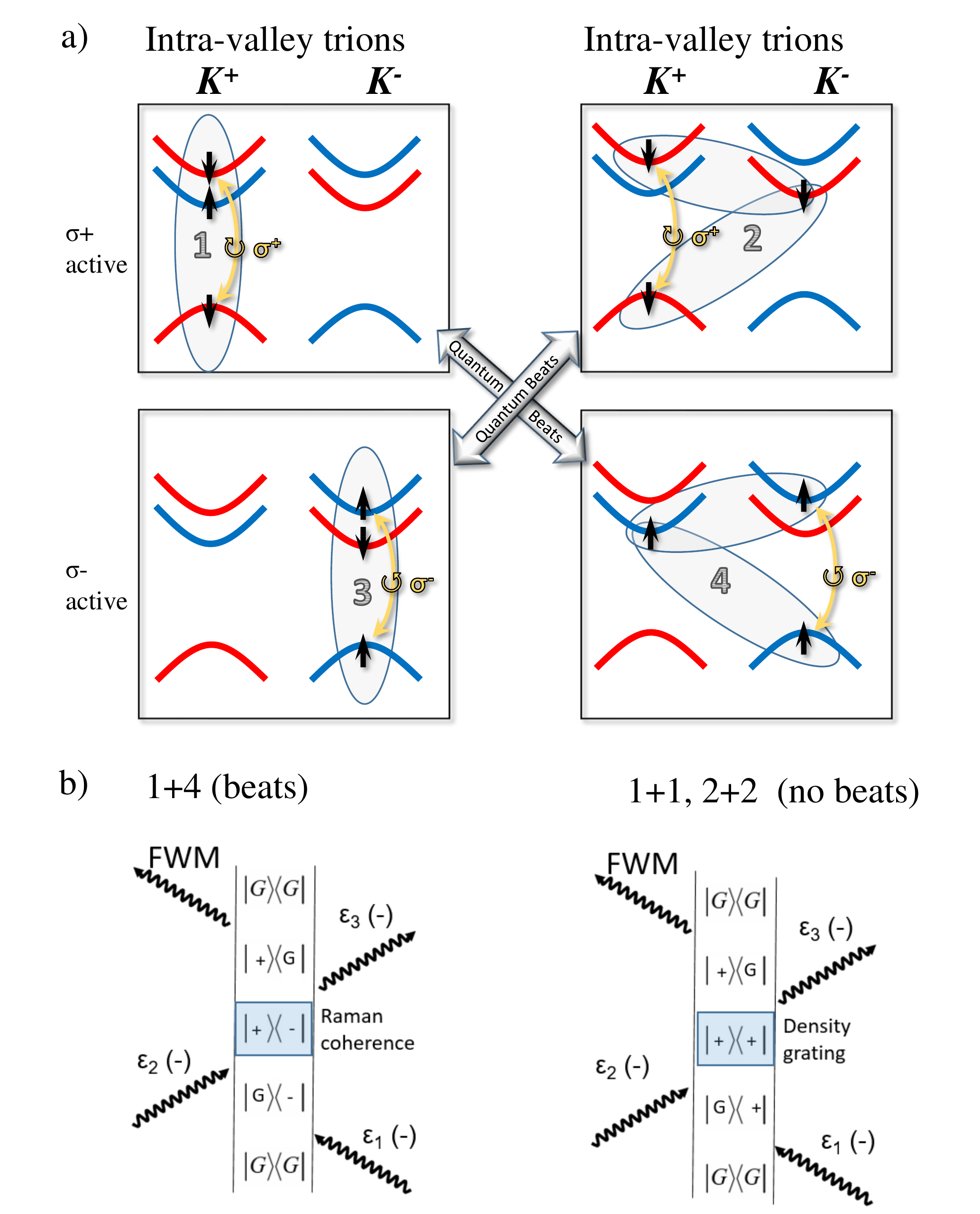}
\caption{\bf Scheme of possible trion electronic configurations and
Feynman diagrams of possible quantum pathways under colinear
excitation (we omit the symmetric diagrams, which can be obtained by
inversing valleys and polarizations to opposite).}
\label{fig:Fig_Scheme_trions}
\end{figure}

\end{document}